# Characterizing Big Data Management

## *Rogério Rossi & Kechi Hirama*
## *University of São Paulo, São Paulo, Brazil*

## rossirogerio@hotmail.com   kechi.hirama@usp.br

## Abstract

Big data management is a reality for an increasing number of organizations in many areas and represents a set of challenges involving big data modeling, storage and retrieval, analysis and visualization. However, technological resources, people and processes are crucial to facilitate the management of big data in any kind of organization, allowing information and knowledge from a large volume of data to support decision-making. Big data management can be supported by these three dimensions: technology, people and processes. Hence, this article discusses these dimensions: the technological dimension that is related to storage, analytics and visualization of big data; the human aspects of big data; and, in addition, the process management dimension that involves in a technological and business approach the aspects of big data management.

**Keywords**: Big Data, Big Data Management, Big Data Challenges, Big Data Analytics, Decision-Making.

## Introduction

Big data refers to the idea that a vast amount of data cannot be treated, processed and analyzed in a simplified way. To Bughin, Chui & Manyika (2010), nowadays, the data are identified in several environments in volumes never seen before, doubling every 18 months as a result of many types of databases as proprietary databases, databases derived from Web communities and from other types of intelligent data assets.

Manyika et al. (2011) consider that big data refers to datasets whose size goes beyond typical databases that can be created, stored, managed and analyzed by existing tools; they also consider the need of creating new technologies for managing big data. It can be seen that several areas currently have data volumes from dozens of terabytes to multiple petabytes (thousands of terabytes).

Fisher, DeLine, Czerwinski & Drucker (2012), however, consider that most often big data refer to the conception that the volume of data cannot be treated, processed and analyzed in a simplified way, requiring much more robust technologies, techniques and people with new skills for managing these large data sets.

As can be seen in Borkar, Carey & Li (2012b), actions related to big data reach various sectors for specific purposes, such as: 1) governments and business tracking contents of several Web social networks to perform sentiment analysis; 2) public sector organizations monitoring health research and various networks to evaluate and to treat epidemics; 3) commercial marketing evaluating the actions of people through social networks in order to understand the behavior of their potential customers.

Borkar, Carey & Li (2012b) argue that the support being offered to organizations considering data-intensive computing, research and analysis, as well as the ability to store data are generating significant challenges to big data management.

For Chen, Chiang & Storey (2012) the big data era has reached several sectors, from government and e-commerce to healthcare organizations. The abundance of data in critical and high social impact sectors requires discussions on data and analytics characteristics. There are some examples of potential areas that deal with these vast amounts of data, such as 1) e-commerce and market intelligence, 2) e-government, 3) science, 4) health insurance, and 5) security.

Grover (2014) points out that the ability to extract knowledge from vast amounts of data that are stored providing opportunities for big data systems can be used and applied to many sectors, such as 1) healthcare, 2) mobile networks, 3) video surveillance, 4) media and entertainment, 5) life sciences, 6) transportation, and 7) study environment.

Russom (2013) considers that sensors spread throughout the world produce outrageous amounts of machine data, highlighting the challenges of capturing and managing these vast amounts of data that are generated continuously in real time and in multi-structured format.

For any of the sectors, some issues must be considered to achieve satisfactory results from big data management. Manyika et al. (2011) consider the following factors as relevant to extract the best results with of big data management: 1) data policies definition, 2) specific technology and techniques; 3) organizational change and talents; 4) data access, and 5) infrastructure.

Considering the application of big data by organizations of various industries the collection and storage of data are observed to be held in proportions unimaginable in the past. Examples can be seen in Bughin, Chui & Manyika (2010), who present some results related to Facebook, which in just two years has quintupled its user database, reaching 500 million users; and Manyika et al. (2011) who point out that, if considering specifically 2010, it appears that global companies exceeded seven exabytes (one exabyte corresponds to one million gigabytes) of data stored.

However, Bughin, Chui & Manyika (2010) show that executives in different sectors are wondering about the difficulties to extract the best results from big data, enabling companies to capitalize the best answers from the abundance of data and enabling them to better manage knowledge to provide capacity for decision-making.

Russom (2013) considers some preliminary difficulties for managing big data: 1) groups of people who are from business or technological areas but do not have adequate skills; 2) inadequate data management infrastructure; and 3) treatment of immature types of data from different sources, such as semi-structured or unstructured data.

"Data are flooding in at rates never seen before" as state Bughin, Chui & Manyika (2010, p.7); thus, the application and use of big data are increasingly and a vast amount of data with varying structures have been used by organizations of many sectors. These data are categorized by Russom (2013) as follow: 1) structured data; 2) complex data (hierarquical or legacy sources); 3) semi-structured data; 4) web-logs; 5) unstructured data; 6) social media data; and 7) machine generated data (sensors, RFID, devices).

Thus, the ability to manage big data, i.e., a high volume of data with the extensive variety of data types that should provide rapid responses is a reality for the organizations that must handle with relevant challenges of big data management as the dimensions related to people, technology, and process management.

The characterization of big data management and the relationship with these three dimensions become the purposes of this article, as regards: 1) highlighting the importance and purpose of big data management; 2) addressing and discussing specific needs involving big data management; 3) discussing several technological, human and process aspects related to big data management; and 4) presenting the difficulties and challenges to analyze the vast amount of data and to visualize the results.

To meet the above objectives, this article is organized as follows: section two presents a theoretical development of big data management; section three considers some works that may be related to this research that also treat the aspects and characteristics of big data management; section four discusses people, process, and technology dimensions involving big data management; section five presents the difficulties and challenges to manage big data; and finally, section six presents the conclusion and proposals for future work.

## Theoretical Development of Big Data

Glatt (2014) presents historical factors concerning the term big data, mentioning that this term has been used since the completion of the 1880 Data Census in the United States. At that moment, with no technology or advanced techniques for data collection and organization, the vast amount of data took seven years to process and finally show results.

Borkar, Carey & Li (2012b) go back to the 1970s to present separate consideration of the term big data, the word 'big' at that time, referring to megabytes, and 'big' over time came to mean gigabytes, evolving to terabytes. Currently, the authors mention that this word related to the term big data refers to petabytes and exabytes.

Bedeley & Iyer (2014) suggest that the term big data was introduced in computing in 2005 to define a large volume of data that traditional data management technologies were not able to manage or to process due to their complexity and volume.

While in the area of computing the term has been employed recently, researchers in other areas are presenting results since 2000. Accordingly to Chen, Chiang & Storey (2012) in a survey pointing to quote the keywords 'business intelligence' 'business analytics' and 'big data', the evolution of the latter is quite relevant given that in 2001 only one study was found referencing the term, and in 2011, 95 were found using the specific term 'big data'.

Luzivan & Meirelles (2014) collaborate with researches into the evolution of the term big data in scientific reports presenting results that show that, in scientific journals, in 2010, 15 reports were found employing this term, and in 2013, 380 scientific reports were found considering the same term.

Bedeley & Iyer (2014) present results across top tier IS journals (journals that occupy the top spot according to the MIS journal rankings) allowing checking that in the field of business, only 16 articles were identified mentioning the term big data.

These results present a quantitative insight into the related research, although the results presented by Bedeley & Iyer (2014) also propose a qualitative view of the technical and scientific reports identified. However, the results demonstrate the need for studies and research in the area, given that needs related to big data management are a reality for an increasing number of organizations. As states Gartner (2012), 85% of the companies' infrastructure will be overloaded by big data until 2015. Moreover, as mentioned by Luzivan & Meirelles (2014), several authors showed a lack of academic studies related to big data under broader and integrative analysis.

According to the definition of the word 'big' from the term big data, Borkar, Carey & Li (2012b) mention that it varies over time, from megabytes (1970s) to exabytes (2014). For Luzivan &

Meirelles (2014), this word belongs to the term big data, and can be seen as a large volume of data in an individualized context and as small volume of data in another; or as large volume of data at a given observed moment and small at another. For Jacobs (2009, p. 40) "what makes most big data *big* is repeated observations over time and/or space".

However, Demchenko, Laat & Membrey (2014) argue that 'big' is not specifically restricted to the volume, but also refers to variables addressing variety, velocity, value, and veracity which make up the Big Data 5V Properties.

The expression, or the full term big data, presents diverse definitions observed in the recent scientific literature. Definitions identified for the term big data are verified in Manyika et al. (2011), in Russon (2013), among other scientific reports. However, this article presents a definition proposed in a draft framework of the NIST (National Institute of Standards and Technology) linked to the US Department of Commerce, which corresponds to: "Big data consists of extensive datasets, primarily in the characteristics of volume, velocity, and/or variety that require a scalable architecture for efficient storage, manipulation, and analysis" (NIST, 2014a, p. 5).

Reflections on big data must be able to effectively meet business competitiveness and support decision-making, which should also be related to the information science and knowledge engineering. Issues that have been considered by Turban, Aronson & Liang (2005) and Laudon & Laudon (2007) for a long time address the elements that integrate the application of information management and knowledge management to business. For Brynjolfsson & McAfee (2012), big data management is responsible for seeking to glean intelligence from data and for translating that into business advantage.

The considerations regarding the definition of big data in a practical and effective manner within an organization may consider the 'Big Data 5V properties' presented by Demchenko, Laat & Membrey (2014). This is a way to set the big data in an organization, considering the '5V properties' that represent: 1) volume, 2) variety, 3) velocity, 4) value, and 5) veracity. It is essential to characterize the environment that can first consider the combination of volume and variety of data to be processed to generate intelligence and competitive advantage for the business.

The definition and clarity of the aspects involving the scenario facing big data enable the organization to align with the specific technologies and techniques that are restricted to big data, and require that it has better control of processes and human resources with specific skills to meet needs related to big data management. Brynjolfsson & McAfee (2012) show that business executives are questioning whether big data is another way to say analytics. This makes explicit the need to define and to clarify the particular aspects for managing big data in a consistent and real way and to meet the expectations.

Currently, organizations are not worried with the question of the need of big data, because it is more than the necessity, it's a reality that should be managed. Big data reflects existing scenarios in multiple-sector organizations. There is a vast amount of data with varied structures, as semi-structured, unstructured or multi-structured data, and there is a necessity to provide quick responses, with the implementation of effective mechanisms for big data management, considering new technologies, organization and process changes, and right people.

# Related Works on Big Data Management

The current situation regarding big data management presents diverse studies involving technological issues, issues dealing with data management, data analysis; there are also studies that link big data to business intelligence or to other consolidated information technology approaches. Usually related studies to manage big data propose two approaches, one that strongly addresses the technological and technical issues to institutionalize and to maintain an infrastructure that considers big data and another that seeks to meet the business goals.

Big data management in this research is not restricted to management based exclusively on information technology, but also considers the involvement of human resources as well as the organizational processes for managing big data.

For Russom (2013), there is a difference between managing big data at a technological level and manage big data in order to support successful business objectives. Hence, the author proposes two relevant questions regarding big data management that correspond to: 1) how effectively does an organization have the technological capacity to manage big data?; and, 2) does the management of big data have the ability to support the business goals?

In fact, information engineering and knowledge management, according to Turban, Aronson & Liang (2005), collaborate with the organizations, increasing their capacity of competitive intelligence and decision-making. In this sense, for business objectives, managing big data becomes essential to provide favorable results from the vast amount of data.

Russom (2013) presents evidence from a survey on a number of North American, European and Asian companies showing that only 3% of the organizations were considered to be at a relatively mature state to manage big data. Most of the organizations participating in the survey (37%) reported to be arguing about it without commitments with the institutionalization of big data. Regarding the expectation of when these organizations expect to have big data in production, the majority (22%) believes that only in three years or more, but 10% of the respondents expect to implement the management of big data within 6 months.

A case study of the banking industry presented by Bedeley & Iyer (2014) discloses that this sector has huge volume of data being generated and processed continuously given to issues related to high competitiveness of the sector and the significant increase in customer database. Other issues that lead to the increased volume of data for the sector are mobile banking and e-banking. This requires that data capture, storage, processing, and analysis strategies, i.e., managing big data should be supported by high technology to provide the best results.

Brynjolfsson & McAfee (2012) suggest that organizations to manage big data should particularly consider five areas: 1) leadership, since the era of big data means not just more data, but the ability to extract results; 2) talent management, considering that the most crucial are the data scientists and professionals with skills to deal with the vast volume of data, organizing large data sets that are not only in structured format; 3) technology, as an important component of the strategy for big data; although the available technology has improved significantly for managing big data, it should be considered novel for many IT departments and integration should be performed; 4) decision-making, reflects the need to maximize cross-functional cooperation between people who manage the data and the people who use them, people who understand business problems must be close to certain data and with people who know effective techniques for extracting the best results; and, 5) company culture, a data-driven organization should cease to be guided solely by hunches and stop using the hippo traditional approaches.

Implementation strategies of big data management actions should be considered for organizations and can be checked at NIST (2014b) and Brynjolfsson & McAfee (2012). NIST (2014b) considers four steps that favor the strategic assessment of big data management: 1) identifying and including stakeholders, 2) identifying potential roadblocks, 3) defining achievable goals, and 4) defining 'finished' and 'success' at the beginning of the project.

For Brynjolfsson & McAfee (2012), some steps can guide the use and application of big data management without huge investments in IT, considering a piecemeal approach to generating capacity for big data management: 1) selection of a business unit to test the actions of big data, considering a team of data scientists, 2) identifying five business opportunities based on big data, considering prototype solution for a given period (approximately five weeks), and 3) implementing an innovation process with four steps - a) experimentation, b) measurement, c) sharing, and d) replication.

NIST (2014b) presents two scales to be considered for organizations related to big data management; the first considers the organizational readiness: 1) no big data, 2) ad hoc, 3) opportunistic, 4) systematic, 5) managed, and 6) optimized; and the second scale that deals with organizational adoption: 1) no adoption, 2) project, 3) program, 4) divisional, 5) cross-divisional, and 6) enterprise.

The characteristics that occur in NIST (2014b) are relevant to provide visibility of the situation in which the organization is concerned about the management of big data, i.e., when the management of big data, in a business and technology approach, is able to provide intelligence to improve competitiveness and decision-making.

## People, Process, and Technology for Managing Big Data

As part of an information system and as the principal component of this type of system, data are collected, qualified, stored and processed by information systems to deliver results that satisfy its users. For Laudon & Laudon (2007), information systems consider three dimensions: people, technology and organization (emphasizing the need for organizational processes). In this sense, these dimensions should be considered for intensive management of big data in the organizations: people, technology, and processes.

To improve competitive advantage and decision-making, organizations consider information a fundamental object. In the information era, and more precisely in the era of digital information, this smart asset becomes increasingly necessary for business survival.

O'Brien and Marakas (2013) argue that information systems have three key business roles: 1) supporting processes and operations, 2) supporting decision making by agents of the organization, and 3) supporting strategies for competitive advantage.

To collaborate in decision-making, information systems must meet some basic requirements, such as the type of support offered, frequency and form of information presentation, format of information, and method of processing information (O'Brien & Marakas, 2013).

The need for accurate, fast and concise information means that this is a costly asset for organizations, however, extremely necessary. Big data is hence also considered an important tool in this scenario where it can be treated as fundamental input to decision-making and competitive advantage.

Fisher et al. (2012) argue that many decision makers, from company executives to government agencies to researchers and scientists, would like to base their decisions and actions on information. Therefore, big data analytics as a new discipline is a workflow that distills terabytes of low-value data, transforming them, in some cases, into a single bit of high-value data.

The ability to generate information, as just a single bit of high-value data, from a large amount of data that present different structures is part of what can determine big data management. And, for successfully managing big data, the three dimensions, technology, processes and people, can favor environments where big data is identified.

Therefore, the characteristics of these three dimensions for managing big data are detailed as follows. In a view that allows understanding the need for big data management as specific technologies and techniques with people with different profiles that are involved in various organizational processes, in business or technology areas.

**People dimension** – people related to big data management need new skills, according to Manyika et al. (2011) there may be limits to innate human ability – to the human sensory and cognitive faculties – to process the data torrent.

There are limitations in human abilities to understand and to consume the vast and varied data set related to big data. The need for new skills is not restricted to those who manage data, to people that manipulate, process and manage the related big data technology environment, but mainly the abilities of users and decision makers should be considered to view an extremely large data set to obtain the necessary information for making important decisions.

For Bughin, Chui & Manyika (2010), using experimentation in big data as essential components for managing decision-making requires new capabilities, as well as organizational and cultural changes.

People involved with big data management currently receive positions with varying titles. Russom (2013) presents the following positions as the three most commonly used to manage big data: 1) Data Architect, 2) Data Analyst, and 3) BI Manager or DW Manager. Although Data Scientist appears as a position related to big data, as a specific professional to handle the management of big data, it is not the mostly considered position by the organizations, being considered for managing big data as well as the Application Developer, Business Analyst, and the System Analyst or System Architect.

However, NIST (2014b) proposes specific actors and roles (Figure 1) for big data management, such as: 1) Data Provider, 2) Data Consumer, 3) Big Data Application Provider, 4) Big Data Framework Provider, and 5) System Orchestrator.

Chen, Chiang & Storey (2012) argue that the United States alone will need between 140,000 and 190,000 professionals with deep analytical skills, as well as 1.5 million managers with data-savvy know-how to analyze big data to make effective decisions. If these proportions are amplified, professionals with this profile will generally have profound relevance in the global technology scenario and business.

For Brynjolfsson & McAfee (2012, p. 65) "big data power does not erase the need for human insight. One of the most critical aspects of big data is its impact on how decisions are made and who gets to make them". The ability of managing big data technologically does not overlap the ability big data gives to the decision maker. It represents important aspects that should be considered by the teams within organizations that manage big data as a backdrop to the real competitive advantage.

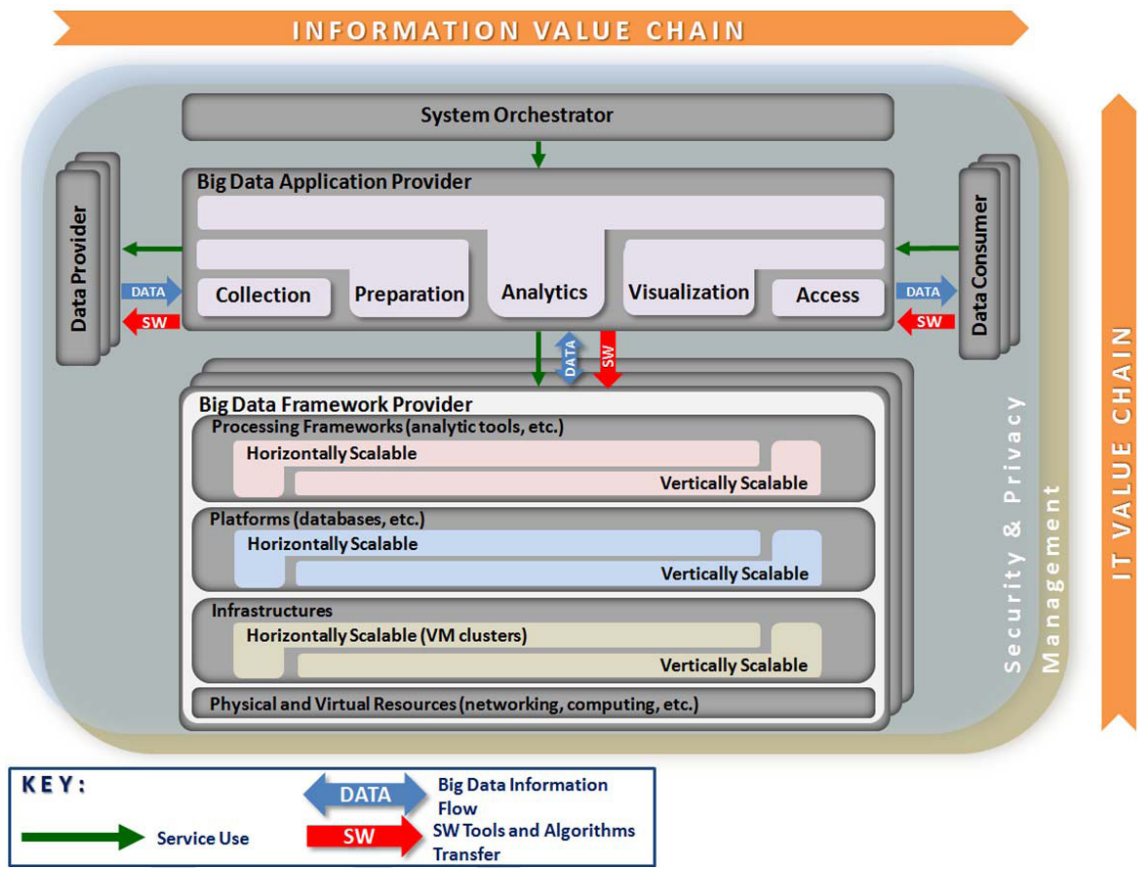

Figure 1: Actors and roles for big data management (NIST, 2014B)

Groups capable of managing big data in an organization, such as data warehouse group, central IT team, or also the business units or department, must possess appropriate skills and pay attention to training programs that promote the proper use to obtain better results from big data management. Russom (2013) proposes 10 top properties for big data management, one of which relates to "get training (and maybe new staff)"; its focus should lie in training and hiring data analysts, data scientists, and data architects who can develop the applications for data exploration, discovery analytics, and real-time monitoring.

Brynjolfsson & McAfee (2012) consider that people who understand the problems need to be together with the right people who manage big data technologies to obtain better results from this vast amount of data.

**Process dimension** – process for big data management are related to the actions that are performed in the technological environment, as in the business environment, i.e., some specific processes should be treated for the technology area where tools are used and specific techniques applied for managing big data; and business processes that are responsible for generating the data, as well as using them accurately after processed.

However, for big data, processes are sometimes interrelated, since it is necessary that they concurrently perform activities related to the business, also performing technical activities, which culminate, for example, in big data analytics.

According to Fisher et al. (2012), there are a number of challenges involving big data and one of them concerns the analysis that must be performed from a vast mass of data that possibly have different structures. For this relevant challenge, the authors present a pipeline that considers a five-step set, representing a data management process, to provide the best results from an analytical visualization.

The pipeline denotes the state of practice for data analysis from a large volume of data. It has been created as the software development waterfall model. The big data pipeline proposed by Fisher et al. (2012) considers the steps that are shown in Figure 2.

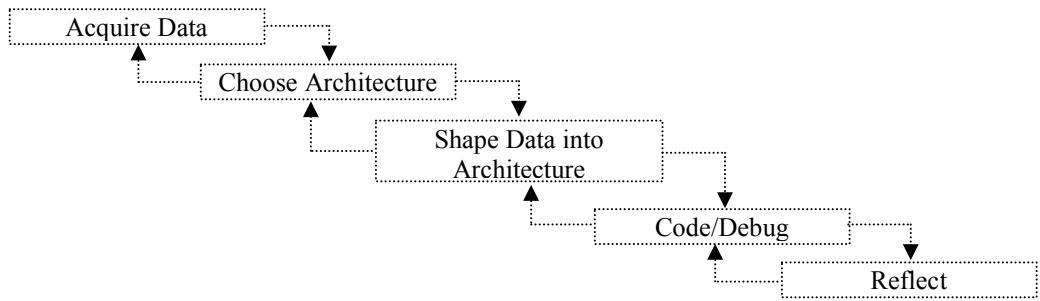

**Figure 2: The Big data Pipeline (Fisher et al., 2012)**

- Acquire data, determines where data are extracted. How to discover the source of data and format relevant subsets to meet the outcomes. Sometimes the data may be stored in schemas that hinder their use. In this case, there are opportunities to improve standards for data storage, streamlining the search and formatting data;
- Choose architecture considers such items as cost and performance. Sometimes the analysis from vast amounts of data requires substantially different abstractions of programming designed for traditional environments. Especially when considering the environment facing Cloud Computing, it imposes nonlinear costs on access, storage and changes what occurs in the environment;
- Shape data into architecture to ensure compatibility when uploading data to the selected platform, a compatible way to computation and data distribution. It is relevant to consider that cloud computing environments use different storage engines from conventional desktops;
- Code / debug suggests the use of specific languages such as R, Python or PIG(data manipulation language) conjugated to Hadoop technology; and
- Reflect corresponds to a step of debugging favoring the visualization and interpretation of results.

Aiming to encourage decision-making, this pipeline can be considered for both, the corporate environment, i.e., the 'business world' to provide answers to the business leaders who still consider techniques such as data mining, machine learning and visualization; and is able to provide answers to scientific research, considering stringent mechanisms for data analysis in which theories and hypotheses could be tested.

For Bizer, Boncz, Brodie & Erling (2011), a five-step methodological overview can serve the needs facing challenges when it comes to the extraction results from 'Big Data World', and this vision includes the following steps:

- Defining the concern – which considers the problem to be solved from the manipulation of data in an environment that considers a vast amount of data;
- Search - search with the vast amount of existing data, i.e., the 'Big Data World', elements that can direct answers to the problem;

- Transform - where the ETL (Extract, Transform, Load) technique is used to perform data extraction from vast amounts of data that are relevant to the solution of the problem, to transform them and to store them for processing;
- Entity resolution - checks the data, ensuring their relevance to the solution of the problem, considering different levels of abstraction; and finally
- Solve the problem - involving actions from the relevant pre-selected data to compute the solution using specific computational domains.

As illustrated, the approaches to manage big data present steps that include business and technology actions, and a joint vision is needed and amplified for people managing big data.

In NIST (2014b), a roadmap to four categories is observed: 1) data services, 2) usage services, 3) capabilities, and 4) vertical orchestrator; as well as the nine features that refer to: 1) storage framework, 2) processing framework, 3) resource managers framework, 4) infrastructure architecture, 5) information architecture, 6) Standard integration framework, 7) application framework, 8) business operations, and 9) business intelligence. These features defined for big data management provide value statements for technology and organizational readiness as presented in Table 1.

Table 1: Big data – technology and organizational features (NIST, 2014b)

| eature | Value statement |
|---|---|
| 1.Storage framework | Define how big data is logically organized, distributed and stored. |
| 2.Processing framework | Define how big data is operated in the big data environment. |
| 3.Resource manager framework | Resource management solutions are required because big data storage and processing frameworks are distributed. |
| 4.Infrastructure architecture | Requires the ability to operate with sufficient network and infrastructure backbone. |
| 5.Information architecture | Data itself needs to be reviewed for its informational value. |
| 6.Standard integration framework | Integration with appropriate standards to assist cross-product integration and knowledge. |
| 7.Application framework | Considering how applications will interact with a big data solution. |
| 8.Business operations | Business operations need to be able to strategize, deploy, and operate big data solution, as big data is more than just technology. |
| 9.Business intelligence | The end value of big data: presenting data as information, intelligence, and insight. |

The value statements defined by NIST (2014b) collaborate with the perception that processes in big data management are integrated, considering business and technology areas.

**Technology dimension** - big data environment considers several technologies and techniques for collecting, storing, processing and analyzing data. Some of these technologies and techniques have emerged specifically in the big data era; other existing ones have been improved for big data. Many techniques and technologies for managing big data have been developed and adapted to add capacity to the analysis that must be performed from big data.

Manyika et al. (2011) present some relevant techniques that consider the statistics and computer science theories and culminating in the big data analytics, among which are:

- Data Mining - technique used to extract patterns from vast amounts of data by combining statistical methods and machine learning data management;
- Genetic Algorithms - technique used for optimization, applied to nonlinear problems;

- Machine Learning - technique that uses artificial intelligence principles and considers the development of algorithms for recognizing complex patterns in large volumes of data and propose intelligent decisions;
- Neural Networks - consider the assumptions of biological neural networks to inspire computational models in identifying patterns in vast amounts of data being used for pattern recognition and optimization.

There are still a number of technologies and techniques in Manyika et al. (2011) that are related to big data environments, which include:

- Hadoop - the open source framework for processing large volumes of data in distributed systems, inspired by tools such as MapReduce and GFS (Google File System) from Google company;
- MapReduce - software framework introduced by Google company to process high volumes of data is also part of the implementation of the Hadoop technology. According to Kulkarni & Khandewal (2014), to address some limitations identified for MapReduce its new generation was proposed, called YARN (Yet Another Resource Negotiator);
- Business Intelligence - refers to a type of application based on software developed to display and to analyze the data; and
- Cloud Computing - technology refers to a computing paradigm with a high level of computational resources sometimes configured as distributed systems to provide services through digital networks.

Agrawal, Das & Abbadi (2011) consider that the field of big data analysis, MapReduce paradigm as well as its open source implementation, Hadoop, corresponds to technologies that have been adopted by the industry as well as by the academia.

Bakshi (2012) also considers the relevance of the related unstructured data, such as texts, electronic messages (e-mails) that use NoSQL (Not Only SQL) database for managing and manipulating unstructured data.

In Borkar, Carey & Li (2012a) and Borkar, Carey & Li (2012b) is identified as the Asterix Project that began at UC Irvine in early 2009 for creating a new parallel, semi-structured information management system. The top layer of Asterix is a DBMS Manager (Data Base Management System) completely parallel with data model (ADM - Asterix Data Model) and data query (AQL - Asterix Query Language) to describe, to analyze and to manipulate data. Asterix software stack is a full parallel DBMS with flexible data model (ADM) and query language (AQL) for describing, querying, and analyzing data.

Hall et al. (2009) present the WEKA (Waikato Environment for Knowledge Analysis) that aims to provide a comprehensive collection of machine learning algorithms and data preprocessing tools to researchers and practitioners. WEKA has specific characteristics for activities of data mining that can be applied to big data environment.

With the techniques and technologies available, the term 'analytics' for big data has been used constantly, and for Fisher et al. (2012) the term analytics is often used broadly to cover any data-driven decision-making. Analytics in the corporate world is considering statistics, data mining, machine learning, and visualization to answer questions that business executives pose. In the academic world, research scientists analyze data sets to form theories and test hypothesis.

The tools for big data analytics, as in Demchenko, Laat & Membrey (2014), are currently offered by major big data technology providers, such as Amazon Elastic MapReduce and Dynamo, Microsoft Azure HDInsight, IBM Big Data Analytics, Cloudera; and others.

Russom (2013) presents some vendors platforms and tools for managing big data, such as:

- Cloudera – as a leader enterprise analytic data management powered by Apache Hadoop;
- Oracle Big Data Appliance –integrates and optimizes all the hardware and software components to build comprehensive analytic application;
- Pentaho – presents Pentaho Data Integration (PDI), an enterprise class, graphical ETL tool; and
- SAP Hana – a smart data access to push queries into Hive Hadoop.

Manyika et al. (2011) mention other technologies and tools that support big data, such as:

- Cassandra – an open source (free) database management system for handing huge amount of data in a distributed system;
- MongoDB - a cross-platform document-oriented database; and
- Dynamo – a proprietary distributed data storage system developed by Amazon.

Besides tools and new technologies, new software languages emerge related to big data environment, such as language 'R', which is an open source programming language for statistical computing and graphics; PIG, a language for data manipulation coupled with Hadoop technology; and AQL (Asterix Query Language) a comparable language such as PIG.

However, Chen, Chiang & Storey (2012) state that data analytics continues to be an active area of research given that statistical machine learning, techniques such as Bayesian networks, Hidden Markov Models, support vector machine, reinforcement learning have been applied to data, text, and web analytics application.

For Brynjolfsson & McAfee (2012), all possible technologies and techniques for big data require a skill set that is new to most IT departments, which will need to work hard to integrate all the relevant internal and external sources of data to increase the decision-making capability based on big data.

## Challenges on Big Data Management

The challenges to big data management refer mainly to structural problems of managing large volumes of data, and especially the difficulties inherent to the ability to extract meaning from this mass of data. For Borkar, Carey & Li (2012b) information has great potential value for many purposes if captured and aggregated effectively.

Big data refers to considerations of data collected and stored in proportions unimaginable in the past. According to Bizer et al. (2011), in the big data world, the databases are unbelievably large in scale, scope, distribution, heterogeneity, and supporting technologies. As examples, according to Dounde (2014), the global data generated from the beginning until the year of 2003 can be estimated to represent about 5 exabytes (one exabyte corresponds to one million gigabytes) and the volume of data generated until 2012 is equal to 2.7 zettabytes (one zettabyte correspond to a thousand exabytes). They are expected to grow by 3 times until 2015.

However, the challenges are not restricted to extracting, storing and managing vast amounts of data, but also refer to semantic analysis of these data, as shown by Bakshi (2012) as it is related to the needs of new skills by technology professionals and users, change in organizational culture and integration environment.

Bizer et al. (2011) consider two classes of primary challenges to big data: 1) engineering - efficiency in data management at unimaginable scales, and 2) semantic - identification of meaning, considering the information that is relevant to specific goals.

Alexander, Hoise and Szalay (2011) also consider that one of the challenges of big data is that data cannot be simply moved or made available for analysis. These should be analyzed *in situ* and/or specific methods must be developed for extracting smaller collections of relevant data to be analyzed and to provide the expected results.

Russom (2013) presents results of a survey that was conducted with professionals of American, European and Asian companies, and points out that for most of them, big data represents an opportunity that enables data exploration and predictive analytics to discover new facts about customers, markets, partners, costs, and operations. A tiny minority considers big data management a problem; although big data poses technical challenges, data volume for a few organizations is a showstopper.

The three most relevant barriers for big data management according to Russom (2013) are: 1) inadequate staffing or skills; 2) lack of governance or stewardship and lack of business sponsorship, and 3) data integration complexity and data ownership and other policies.

Borkar, Carey & Li (2012b) consider that big data analytics and management is being touted as a critical challenge in the current computing landscape given that governments and businesses are tracking the content of blogs and tweets to perform sentiment analysis; likewise, health insurance organizations are monitoring search trends to check the progress of epidemics. Social scientists are also using social information from different social networks to be used more effectively for the public good.

Grover (2014) considers that big data environments create significant opportunities and challenges. Specifically, technology organizations must find ways to cope with security and other technical challenges for managing the massive volumes of data, such as: 1) heterogeneity and incompleteness of data, 2) scale (large and rapidly increasing volumes of data), 3) context awareness, 4) performance issues, 5) security and privacy, and 6) other challenges (timeliness of data analysis, distributed storage structures, content validation, stream processing and real time analytics).

Brynjolfsson & McAfee (2012) show that the five large areas for big data management (leadership, talent management, technology, decision-making, and company culture) still reflect challenges to organizations, i.e., the talents and organizational changes should be revised to encourage the big data world; technologies that are extremely relevant, even though they are being offered, they are sometimes not easily integrated into the environment. Moreover, there is the need for greater cooperation and integration among people who understand the problems and master the technologies to promote support to decision-making.

Opportunities and challenges can also be seen in Luzivan & Meirelles (2014), related to studies and research on big data. The authors join others to express that there is a significant need for participation by academic researchers in overall amplitude. They also highlights gaps in the field of Information Science related to big data, and feature five groups of issues that may inspire research related to big data: 1) multidisciplinary studies, 2) methodological standards, 3) structure, 4) ethics, security and access, and 5) human capital formation.

## Conclusion and Further Works

Big data management is a reality for many types of organizations and poses a challenge to computer science and information technology. The primary characteristics of big data are related to its volume and variety as also other characteristics have been considered as velocity, value, and veracity.

Big data influences public and private sectors, science and economy, areas such as education and healthcare, among others. The proposal for big data scenario features important actions related to the provision of new technologies and techniques and its integration with existing technologies to promote the expected results. The specific needs related to big data involve the ability to manage the infrastructure and its semantic capacity, i.e., the ability related to improve decision-making.

Big data management reflects several aspects, both in the technological field and in how to manage teams (technical or users); big data management is also strongly related to the processes involved that consider aspects related to collection, storage, and retrieval of data by using technical and user knowledge.

Thus, human, technological dimensions and related processes for big data management are able to provide more favorable conditions to this new scenario. However, the three dimensions require further studies and researches. Even though the technological conditions for big data management have evolved, they are not enough by themselves, because they need to be integrated and operated by qualified personnel. Hence, the demand for new skills and trained professionals to use the new technological arsenal emerges.

In addition, the processes must be more effective, sometimes requiring major revisions, either in the user's view, or from the technology point of view. Both need to review their work processes to provide better results from the vast amounts of data presented on a daily basis. The process affects how to generate, store, and retrieve data, and its presentation and visualization.

The difficulties and challenges in the field of engineering and the semantic value that can be provided by big data reflect numerous future works in this area. As engineering issues, the necessary infrastructure for big data environment and its integration is presented as a fundamental aspect. Moreover, human issues must be addressed to operate big data; the semantic value extracted from big data requires more robust processes, processes from business or tech processes. Analysis and visualization are both especially critical processes for big data management.

## Biographies

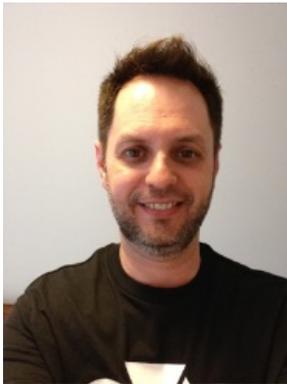

**Rogério Rossi** received his B. S. in Mathematics by the University Center Foundation Santo André as he also has a M.S. and Ph.D. in Electrical Engineering, both by Mackenzie Presbyterian University. He is in a Postdoctoral Program at the University of São Paulo developing researches that are related to Complex Systems, Big Data and the Internet of Things (IoT).

He is an Adjunct Professor for Information Technology and Computer Science courses of graduate and undergraduate programs in São Paulo. He has done research on the fields of software quality, and quality for digital educational solutions and he also has some publications on this area.

He is a member of IACSIT (International Association of Computer Science and Information Technology) and he worked as a reviewer for InSite Conferences'2013 and 2015, and e-Skills Conference'2014; as he also presented his papers in the InSite Conferences in Montreal, Canada (2012) and Porto, Portugal (2013).

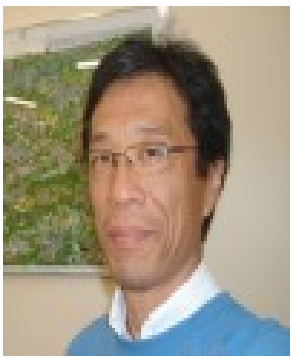

Kechi Hirama received his B.S., M.S., Ph.D. and Associate Professor degrees in Computer Engineering from Escola Politécnica of the University of São Paulo, São Paulo, Brazil in 1980, 1988, 1996 and 2008, respectively.

He worked 15 years in the Control and Automation area in research organizations and since 1996 he has been a Professor of the Department of Computer and Digital Systems Engineering of Escola Politécnica of the University of São Paulo.

His interests include Complex System, System Dynamics, Big Data and Internet of Things (IoT).